\documentstyle[12pt]{article}

\input amssym.tex

\begin{document}

%%%%%%%%%%%%%%%%%%%%%%%%%%%%%%%%%%%%%%%%%%%%%%%%%%%%%%%%%%%%%
%%%%%%%%%%%%%%%%%%%%%%%%%%%%%%%%%%%%%%%%%%%%%%%%%%%%%%%%%%%%
%%%%%%%%%%%%%%%%%%%%%%%%%%%%%%%%%%%%%%%%%%%%%%%%%%%%%%%%%%%%%%%%%%
%%%%  MACROS:
%**********************************************************************
%GREEK LETTERS
\renewcommand{\a}{\alpha}
\renewcommand{\b}{\beta}
\newcommand{\g}{\gamma}           \newcommand{\G}{\Gamma}
\renewcommand{\d}{\delta}         \newcommand{\D}{\Delta}
\newcommand{\ve}{\varepsilon}
\newcommand{\eps}{\epsilon}
\newcommand{\k}{\kappa}
\newcommand{\ld}{\lambda}        \newcommand{\LD}{\Lambda}
\newcommand{\om}{\omega}         \newcommand{\OM}{\Omega}
\newcommand{\p}{\psi}             \newcommand{\PS}{\Psi}
\newcommand{\ro}{\rho}
\newcommand{\s}{\sigma}           \renewcommand{\S}{\Sigma}
\newcommand{\th}{\theta}         \newcommand{\T}{\Theta}
\newcommand{\f}{{\phi}}           \newcommand{\F}{{\Phi}}
\newcommand{\vf}{{\varphi}}
\newcommand{\y}{{\upsilon}}       \newcommand{\Y}{{\Upsilon}}
\newcommand{\z}{\zeta}
\newcommand{\X}{\Xi}
%************************************************************************
%  CAL. LETTERS
\newcommand{\cA}{{\cal A}}
\newcommand{\cB}{{\cal B}}
\newcommand{\cC}{{\cal C}}
\newcommand{\cD}{{\cal D}}
\newcommand{\cE}{{\cal E}}
\newcommand{\cF}{{\cal F}}
\newcommand{\cG}{{\cal G}}
\newcommand{\cH}{{\cal H}}
\newcommand{\cI}{{\cal I}}
\newcommand{\cJ}{{\cal J}}
\newcommand{\cK}{{\cal K}}
\newcommand{\cL}{{\cal L}}
\newcommand{\cM}{{\cal M}}
\newcommand{\cN}{{\cal N}}
\newcommand{\cO}{{\cal O}}
\newcommand{\cP}{{\cal P}}
\newcommand{\cQ}{{\cal Q}}
\newcommand{\cS}{{\cal S}}
\newcommand{\cR}{{\cal R}}
\newcommand{\cT}{{\cal T}}
\newcommand{\cU}{{\cal U}}
\newcommand{\cV}{{\cal V}}
\newcommand{\cW}{{\cal W}}
\newcommand{\cX}{{\cal X}}
\newcommand{\cY}{{\cal Y}}
\newcommand{\cZ}{{\cal Z}}
%***************************************************************
% CAPITAL LETTERS WITH HAT
\newcommand{\hA}{{\widehat A}}
\newcommand{\hB}{{\widehat B}}
\newcommand{\hC}{{\widehat C}}
\newcommand{\hD}{{\widehat D}}
\newcommand{\hE}{{\widehat E}}
\newcommand{\hF}{{\widehat F}}
\newcommand{\hG}{{\widehat G}}
\newcommand{\hH}{{\widehat H}}
\newcommand{\hI}{{\widehat I}}
\newcommand{\hJ}{{\widehat J}}
\newcommand{\hK}{{\widehat K}}
\newcommand{\hL}{{\widehat L}}
\newcommand{\hM}{{\widehat M}}
\newcommand{\hN}{{\widehat N}}
\newcommand{\hO}{{\widehat O}}
\newcommand{\hP}{{\widehat P}}
\newcommand{\hQ}{{\widehat Q}}
\newcommand{\hS}{{\widehat S}}
\newcommand{\hR}{{\widehat R}}
\newcommand{\hT}{{\widehat T}}
\newcommand{\hU}{{\widehat U}}
\newcommand{\hV}{{\widehat V}}
\newcommand{\hW}{{\widehat W}}
\newcommand{\hX}{{\widehat X}}
\newcommand{\hY}{{\widehat Y}}
\newcommand{\hZ}{{\widehat Z}}
%**********************************************************************
% LETTERS WITH HAT
\newcommand{\Ha}{{\widehat a}}
\newcommand{\Hb}{{\widehat b}}
\newcommand{\Hc}{{\widehat c}}
\newcommand{\Hd}{{\widehat d}}
\newcommand{\He}{{\widehat e}}
\newcommand{\Hf}{{\widehat f}}
\newcommand{\Hg}{{\widehat g}}
\newcommand{\Hh}{{\widehat h}}
\newcommand{\Hi}{{\widehat i}}
\newcommand{\Hj}{{\widehat j}}
\newcommand{\Hk}{{\widehat k}}
\newcommand{\Hl}{{\widehat l}}
\newcommand{\Hm}{{\widehat m}}
\newcommand{\Hn}{{\widehat n}}
\newcommand{\Ho}{{\widehat o}}
\newcommand{\Hp}{{\widehat p}}
\newcommand{\Hq}{{\widehat q}}
\newcommand{\Hs}{{\widehat s}}
\newcommand{\Hr}{{\widehat r}}
\newcommand{\Ht}{{\widehat t}}
\newcommand{\Hu}{{\widehat u}}
\newcommand{\Hv}{{\widehat v}}
\newcommand{\Hw}{{\widehat w}}
\newcommand{\Hx}{{\widehat x}}
\newcommand{\Hy}{{\widehat y}}
\newcommand{\Hz}{{\widehat z}}
%%%%%%%%%%%%%%%%%%%%%%%%%%%%%%%%%%%%%%%%%%%%%%%%%%%%%
%%%%%%%%%%%%%%%%%%%%%%%%%%%%%%%%%%%%%%%%%%%%%%%%%%%%%%%
%SPECIAL CHARACTERS
\newcommand{\deff}{\,\stackrel{\rm def}{\equiv}\,}
\newcommand{\lra}{\longrightarrow}
\newcommand{\ra}{\,\rightarrow\,}
\def\limar#1#2{\,\raise0.3ex\hbox{$\longrightarrow$\kern-1.5em\raise-1.1ex
\hbox{$\scriptstyle{#1\rightarrow #2}$}}\,}
\def\limarr#1#2{\,\raise0.3ex\hbox{$\longrightarrow$\kern-1.5em\raise-1.3ex
\hbox{$\scriptstyle{#1\rightarrow #2}$}}\,}
\def\limlar#1#2{\ \raise0.3ex
\hbox{$-\hspace{-0.5em}-\hspace{-0.5em}-\hspace{-0.5em}
\longrightarrow$\kern-2.7em\raise-1.1ex
\hbox{$\scriptstyle{#1\rightarrow #2}$}}\ \ }
\newcommand{\limm}[2]{\lim_{\stackrel{\scriptstyle #1}{\scriptstyle #2}}}
\newcommand{\wt}{\widetilde}
\newcommand{\os}{{\otimes}}
\newcommand{\da}{{\dagger}}
\newcommand{\stimes}{\times\hspace{-1.1 em}\supset}
\def\h{\hbar}
\newcommand{\ih}{\frac{\i}{\h}}
\newcommand{\exx}[1]{\exp\left\{ {#1}\right\}}
\newcommand{\ord}[1]{\mbox{\boldmath{$\cO$}}\left({#1}\right)}
%%%%%%%%%%%%%%%%%%
%% OPERATOR UNITY (modified 1)
\newcommand{\one}{{\leavevmode{\rm 1\mkern -5.4mu I}}}
%%%%%%%%%%%%%%%%%%
%% SETS OF NUMBERS
%    modified  Z                meaning:  integers
\newcommand{\Z}{Z\!\!\!Z}
%
%%%%%%%%%%%%%%%%%%
% The auxiliary definitions
\newcommand{\Ibb}[1]{ {\rm I\ifmmode\mkern
            -3.6mu\else\kern -.2em\fi#1}}
\newcommand{\ibb}[1]{\leavevmode\hbox{\kern.3em\vrule
     height 1.2ex depth -.3ex width .2pt\kern-.3em\rm#1}}
% are only used here:
%%%%%%%%%%%%%%%%%%%%%
%    modified   N               meaning:   natural numbers
\newcommand{\N}{{\Ibb N}}
%    modified   C               meaning:   complex numbers
\newcommand{\C}{{\ibb C}}
%    modified   R               meaning:   real numbers
\newcommand{\R}{{\Ibb R}}
%    modified   H               meaning:   quaternions
\newcommand{\HH}{{\Ibb H}}
\newcommand{\rational}{{\kern .1em {\raise .47ex
\hbox{$\scripscriptstyle |$}}
    \kern -.35em {\rm Q}}}
%%%%%%%%%%%%%%%%%%%%%%%%
%%%%%%%%%%%%%%%%%%%%%%%%%%%%%%%%%%%%%%%%
\newcommand{\bm}[1]{\mbox{\boldmath${#1}$}}
\newcommand{\intf}{\int_{-\infty}^{\infty}\,}
%%%%%% HILBERT SPACES %%%%%%%%%%%%%%%%
\newcommand{\LL}{\cL^2(\R^2)}
\newcommand{\LLS}{\cL^2(S)}
%%%%%%%%%%%%%%%%%%%%%%%%%%%%%%%%%%%%%%%
%%%%%% REAL AND IMAGINARY PARTS %%%%%%%%
\newcommand{\Ree}{{\cal R}\!e \,}
\newcommand{\Imm}{{\cal I}\!m \,}
%%%%%%%%%%%%%%%%%%%%%%%%%%%%%%%%%%%%%%%
%% SOME SPECIAL SYMBOLS IN ROMAN
\newcommand{\tr}{{\rm {Tr} \,}}
\newcommand{\er}{{\rm{e}}}
\renewcommand{\i}{{\rm{i}}}
\newcommand{\divv}{{\rm {div} \,}}
\newcommand{\id}{{\rm{id}\,}}
\newcommand{\ad}{{\rm{ad}\,}}
\newcommand{\Ad}{{\rm{Ad}\,}}
\newcommand{\const}{{\rm{\,const\,}}}
\newcommand{\rank}{{\rm{\,rank\,}}}
\newcommand{\diag}{{\rm{\,diag\,}}}
\newcommand{\sign}{{\rm{\,sign\,}}}
%%%%%%%%%%%%%%%%%%%%%%%%%%%%%%%%%%%%%%%%%%%%%%%%%%%%%%%%
%%%DERIVATIVES%%%
\newcommand{\pa}{\partial}
\newcommand{\pad}[2]{{\frac{\partial #1}{\partial #2}}}
\newcommand{\padd}[2]{{\frac{\partial^2 #1}{\partial {#2}^2}}}
\newcommand{\paddd}[3]{{\frac{\partial^2 #1}{\partial {#2}\partial {#3}}}}
\newcommand{\der}[2]{{\frac{{\rm d} #1}{{\rm d} #2}}}
\newcommand{\derr}[2]{{\frac{{\rm d}^2 #1}{{\rm d} {#2}^2}}}
\newcommand{\fud}[2]{{\frac{\delta #1}{\delta #2}}}
\newcommand{\fudd}[2]{{\frac{\d^2 #1}{\d {#2}^2}}}
\newcommand{\fuddd}[3]{{\frac{\d^2 #1}{\d {#2}\d {#3}}}}
\newcommand{\dpad}[2]{{\displaystyle{\frac{\partial #1}{\partial #2}}}}
\newcommand{\dfud}[2]{{\displaystyle{\frac{\delta #1}{\delta #2}}}}
\newcommand{\dd}{\partial^{(\ve)}}
\newcommand{\ddd}{\bar{\partial}^{(\ve)}}
\newcommand{\dfrac}[2]{{\displaystyle{\frac{#1}{#2}}}}
\newcommand{\dsum}[2]{\displaystyle{\sum_{#1}^{#2}}}
\newcommand{\dint}{\displaystyle{\int}}
\newcommand{\dg}{\!\not\!\partial}
\newcommand{\vg}[1]{\!\not\!#1}
%%%%%%%%%%%%%%%%%%%%%%%%%%%%%%%%%%%%%%%%%
%%%%%%%%%%%%%%%%%%%%%%%%%%%%%%%%%%%%%%%%%%%
%%%% BRA and KET                   %%%%%%%
\def\<{\langle}
\def\>{\rangle}
\def\lgl{\langle\langle}
\def\rgr{\rangle\rangle}
\newcommand{\bra}[1]{\left\langle {#1}\right|}
\newcommand{\ket}[1]{\left| {#1}\right\rangle}
\newcommand{\vev}[1]{\left\langle {#1}\right\rangle}
%%%%%%%%%%%%%%%%%%%%%%%%%%%%%%%%%%%%%%%%%%%%
%%%%%    EQUATIONS          %%%%%%%%%%%%%%
\newcommand{\be}{\begin{equation}}
\newcommand{\ee}{\end{equation}}
\newcommand{\bn}{\begin{eqnarray}}
\newcommand{\en}{\end{eqnarray}}
\newcommand{\bnn}{\begin{eqnarray*}}
\newcommand{\enn}{\end{eqnarray*}}
\newcommand{\e}{\label}
\newcommand{\nbr}{\nonumber\\[2mm]}
\newcommand{\r}[1]{(\ref{#1})}
\newcommand{\refp}[1]{\ref{#1}, page~\pageref{#1}}
\renewcommand {\theequation}{\thesection.\arabic{equation}}
\renewcommand {\thefootnote}{\fnsymbol{footnote}}
%%%%%%%%%%%%%%%%%%%%%%%%%%%%%%%%%%%%%%%%
%%% SPACES      %%%%%%%%%%%%%%%%%%%%%%%%%%%
\newcommand{\qq}{\qquad}
\newcommand{\qqq}{\quad\quad}
%%%%%%%%%%%%%%%%%%%%%%%%%%%%%%%%%%%%%%%%%%%%%%%%
%%%%%%%    ADDITIONAL COMMANDS   %%%%%%%%%%%%%
\newcommand{\biz}{\begin{itemize}}
\newcommand{\eiz}{\end{itemize}}
\newcommand{\ben}{\begin{enumerate}}
\newcommand{\een}{\end{enumerate}}
%%%%%%%%%%%%%%%%%%%%%%%%%%%%%%%%%%%%%%%%%%%%%%%%%%%%%%%%%%%%%%%
\def\nc{noncommutative }
\def\ncy{noncommutativity }
\def\com{commutative }
\def\JLD{Jost-Lehmann-Dyson }
\def\th{$\theta_{\mu\nu}$}
\def \simlt{\stackrel{<}{{}_\sim}}
\def\ss{$\theta_{0i}=0$}
\def\P{Poincar\'e}
%%%%%%%%%%%%%%%%%%%%%%%%%%%%%%%%%%%%%%%%%%%%%%%%%%%%%%%%%%%%%%
\thispagestyle{empty}
\begin{flushright}
%HIP-2004-??/TH\\
%hep-th/0402??? %\vskip 14mm
\end{flushright}

\begin{center}

{\large{\bf{Analyticity of the Scattering Amplitude,\\Causality
and High-Energy Bounds\\
in Quantum Field Theory on Noncommutative Space-Time\\
 }}} \vskip
.7cm

{\bf{\large{Anca Tureanu}}

{\it High Energy Physics Division, Department of Physical
Sciences,
University of Helsinki\\
\ \ {and}\\
\ \ Helsinki Institute of Physics, P.O. Box 64, FIN-00014
Helsinki, Finland}}

\end{center}

\setcounter{footnote}{0}

{\bf Abstract} In the framework of quantum field theory (QFT) on
noncommutative (NC) space-time with the symmetry group $O(1,1)\times
SO(2)$, we prove that the Jost-Lehmann-Dyson representation, based
on the causality condition taken in connection with this symmetry,
leads to the mere impossibility of drawing any conclusion on the
analyticity of the $2\rightarrow 2$-scattering amplitude in
$\cos\Theta$, $\Theta$ being the scattering angle. Discussions on
the possible ways of obtaining high-energy bounds analogous to the
Froissart-Martin bound on the total cross-section are also
presented.

%\vspace*{0.1cm}
%\begin{multicols} {2}

\newpage
\section{Introduction}

The development of QFT on NC space-time, especially after the
seminal work of Seiberg and Witten \cite{SW}, which showed that the
NC QFT arises from string theory, has triggered lately the interest
also towards the formulation of an axiomatic approach to the
subject. The power of the axiomatic approach consists in that that
the results are rigorously derived, with no reference to the
specific form of interaction or to perturbation theory.
Consequently, in the framework of noncommutative spaces, the
analytical properties of scattering amplitude in energy $E$ and
forward dispersion relations have been considered \cite{Liao, CMTV},
Wightman functions have been introduced and the CPT theorem has been
proven \cite{AG, CPT}, and as well attempts towards a proof of the
spin-statistics theorem have been made \cite{CPT}\footnote {In the
context of the Lagrangean approach to NC QFT, the CPT and
spin-statistics theorems have been proven in general in \cite{CNT};
for CPT invariance in NC QED, see \cite{Shahin,Kostelecky}, and in
NC Standard Model \cite{Aschieri}.}.

In the axiomatic approach to commutative QFT, one of the fundamental
results consisted of the rigorous proof of the Froissart bound on
the high-energy behaviour of the scattering amplitude, based on its
analyticity properties \cite{Froissart, Martin}. In this paper we
aim at obtaining the analog of this bound when the space-time is
noncommutative. Such an undertaking, besides being topical in
itself, would also prove fruitful in the conceptual understanding of
subtle issues, such as causality, in nonlocal theories to which the
NC QFT's belong.

In the following we shall consider NC QFT on a space-time with the
commutation relation
\be\label{cr}[x_\mu,x_\nu]=i\theta_{\mu\nu}\ , \ee
where $\theta_{\mu\nu}$ is an antisymmetric constant matrix (for a
review, see, e.g., \cite{DN,Szabo}). Such NC theories violate
Lorentz invariance, while translational invariance still holds. We
can always choose the system of coordinates, such that
$\theta_{13}=\theta_{23}=0$ and
$\theta_{12}=-\theta_{21}\equiv\theta$. Then, for the particular
case of space-space noncommutativity, i.e. $\theta_{0i}=0$, the
theory is invariant under the subgroup $O(1,1)\times SO(2)$ of the
Lorentz group. The requirement that time be commutative
($\theta_{0i}=0$) discards the well-known problems with the
unitarity \cite{unit} of the NC theories and with causality
\cite{causal1,causal2} (see also \cite{CNT}). As well, the
$\theta_{0i}=0$ case allows a proper definition of the $S$-matrix
\cite{CMTV}.

In the conventional (commutative) QFT, the Froissart bound was
first obtained \cite{Froissart} using the conjectured Mandelstam
representation (double dispersion relation) \cite{mandel}, which
assumes analyticity in the entire $E$ and $\cos\Theta$ complex
planes. The Froissart bound,
\be\label{fm} \sigma_{tot}(E)\leq c\ \ \ln^2 \frac{E}{E_0}\ , \ee
expresses the upper limit of the total cross-section
$\sigma_{tot}$ as a function of the CMS energy $E$, when
$E\to\infty$. However, such an analyticity or equivalently the
double dispersion relation has not been proven, while smaller
domains of analyticity in $\cos\Theta$ were already known
\cite{Lehmann}.

One of the main ingredients in rigorously obtaining the Froissart
bound is the Jost-Lehmann-Dyson representation \cite{JL,D} of the
Fourier transform of the matrix element of the commutator of
currents, which is based on the causality as well as the spectral
conditions (for an overall review, see \cite{Schweber}). Based on
this integral representation, one obtains the domain of analyticity
of the scattering amplitude in $\cos\Theta$. This domain proves to
be an ellipse $-$ the so-called Lehmann's ellipse \cite{Lehmann}.

However, this domain of analyticity in $\cos\Theta$ can be enlarged
to the so-called Martin's ellipse by using the dispersion relations
satisfied by the scattering amplitude and the unitarity constraint
on the partial-wave amplitudes. Using this larger domain of
analyticity, the Froissart bound (\ref{fm}) was rigorously proven in
QFT \cite{Martin} (for a review, see \cite{review}).

Further on, the analog of the Froissart-Martin bound was
rigorously obtained for the $2 \to 2$-particle scattering in a
space-time of arbitrary dimension $D$ \cite{Fischer,CFV}.

In NC QFT with $\theta_{0i}=0$ we shall follow the same path for the
derivation of the high-energy bound on the scattering amplitude,
starting from the Jost-Lehmann-Dyson representation and adapting the
derivation to the new symmetry $O(1,1)\times SO(2)$ and to the
nonlocality of the NC theory\footnote{A preliminary work along this
line with stronger claims, based on a conjecture, has been
previously reported in \cite{JLD}.}. In Section 2 we derive the
Jost-Lehmann-Dyson representation satisfying the light-wedge
(instead of light-cone) causality condition, inspired by the above
symmetry. We show that no analyticity of the scattering amplitude in
$\cos\Theta$ can be obtained in such a case. Since the causality
condition is the key ingredient for the analyticity of the
scattering amplitude, in Section 3 we discuss possible causality
postulates in the noncommutative case, in relation both with the
maximal symmetry of the theory (twisted Poincar\'e \cite{CKT}) and
with the scale of nonlocality as obtained so far in perturbative
calculations. It turns out that by postulating a {\it finite range
of nonlocality}, compatible with the twisted Poincar\'e symmetry,
and by using the global nature of local commutativity, we can obtain
from the Jost-Lehmann-Dyson representation a domain of analyticity
in $\cos\Theta$, which coincides with the Lehmann ellipse. Further,
the extension of this analyticity domain to Martin's ellipse is
possible in the case of the incoming particles' momenta orthogonal
to the NC plane $(x_1,x_2)$, which eventually enables us to derive
the analog of the Froissart-Martin bound (\ref{fm}) for the total
cross-section. The general configuration of incoming particles'
momenta is also discussed, together with the problems which arise in
such a case. However, the perturbative calculations performed so far
seem to indicate an {\it infinite range of nonlocality}, in which
case the initial causality condition involving the light-wedge
should be postulated, leading to the lack of analyticity of the
scattering amplitude. The situation is discussed in connection with
the perturbative problem of UV/IR mixing in NC QFT. Section 6 is
devoted to conclusion and discussions.

\setcounter{equation}{0}

\section{Jost-Lehmann-Dyson representation}

The Jost-Lehmann-Dyson representation \cite{JL,D} is the integral
representation for the Fourier transform of the matrix element of
the commutator of currents:
\be\label{fq} f(q)=\int d^4xe^{iqx}f(x)\ , \ee
where
\be\label{fx} f(x)=\langle
p'|[j_1(\frac{x}{2}),j_2(-\frac{x}{2})]|p\rangle\ ,\ee
satisfying the causality and spectral conditions. The process
considered is the $2\to 2$  scalar particles scattering, $k+p\to
k'+p'$, and $j_1$ and $j_2$ are the scalar currents corresponding
to the incoming and outgoing particles with momenta $k$ and $k'$
(see also \cite {Schweber,BS}).

For NC QFT with $O(1,1)\times SO(2)$ symmetry, in \cite{LAG} a new
causality condition was proposed, involving (instead of the
light-cone) the light-wedge corresponding to the coordinates $x_0$
and $x_3$, which form a two-dimensional space with the $O(1,1)$
symmetry. Accordingly we shall require the vanishing of the
commutator of two currents (in general, observables) at space-like
separations in the sense of $O(1,1)$ as:
\be\label{causal} [j_1(\frac{x}{2}),j_2(-\frac{x}{2})]=0 \ ,\ \
\mbox{for}\ \ \tilde x^2\equiv x_0^2-x_3^2<0\ . \ee

The spectral condition compatible with (\ref{causal}) would
require now that the physical momenta be in the forward
light-wedge:
\be\label{spectr} \tilde p^2\equiv p_0^2-p_3^2>0\ \ \mbox{and}\ \
p_0>0\ . \ee
The standard spectral condition
$$
p_0^2-p_1^2-p_2^2-p_3^2\geq 0,\ \
\ p_0>0\ .
$$
based on Poincar\'e symmetry or twisted Poincar\'e symmetry
\cite{CKT} implies the forward light-wedge condition (\ref{spectr})
as well.

The spectral condition (\ref{spectr}) will impose restrictions on
$f(q)$. Using the translational invariance in (\ref{fx}), one can
express the matrix element of the commutator of currents, $f(x)$,
in the form:
\bn\label{tr_inv} f(x)&=&\int dq
e^{-iqx+i(p+p')\frac{x}{2}}G_1(q)-\int dq
e^{iqx-i(p+p')\frac{x}{2}}G_2(q)\cr &=&\int dq
e^{-iqx}\left[G_1\left(q+\frac{1}{2}(p+p')\right)-G_2\left(-q+\frac{1}{2}(p+p')\right)\right]\
, \en
where
\bn\label{gs}
 G_1(q)&=&\langle p'|j_1(0)|q\rangle\langle
q|j_2(0)|p\rangle\ ,\cr G_2(q)&=&\langle p'|j_2(0)|q\rangle\langle
q|j_1(0)|p\rangle\ . \en
Comparing (\ref{tr_inv}) with the inverse Fourier
transformation\footnote{Throughout the paper we omit all the
inessential factors of $(2\pi)^n$, which are irrelevant for the
analyticity considerations.}, $f(x)=\int dq e^{-iqx}f(q)$, it
follows that
\be f(q)=f_1(q)-f_2(q)=
G_1\left(q+\frac{1}{2}(p+p')\right)-G_2\left(-q+\frac{1}{2}(p+p')\right)\
.\ee
Given the way the functions $G_1$ and $G_2$ are defined in
(\ref{gs}), one finds that  $f(q)=0$ in the region where the
momenta $q+\frac{1}{2}(p+p')$ and $-q+\frac{1}{2}(p+p')$ are
simultaneously nonphysical, i.e. when they are out of the future
light-wedge (\ref{spectr}).

In order to express the condition for $f(q)=0$, we shall define the
$O(1,1)$-invariant $\tilde m^2=k^2_0-k^2_3=f(m^2, k_1^2+k_2^2)$,
where $k$ is the momentum of an arbitrary state and $m$ is its mass.
However, we have to point out that $\tilde m$ is only a kinematical
variable, invariant with respect to $O(1,1)$ (but not the mass).

For the {\it physical} states with momentum $q+\frac{1}{2}(p+p')$,
we take $\tilde m_{1}$ to be the minimal value of the
$O(1,1)$-invariant quantity above. Then, in the Breit frame, where
$\frac{1}{2}(p+p')=(p_0,0,0,0)$, one finds that $f_1(q)\neq 0$ for
all the $q$ values, satisfying the spectral condition $q_0+p_0\geq
0$ and $(q_0-p_0)^2-q_3^2\geq 0$. In other words, $f_1(q)=0$ for
$q_0<-p_0+\sqrt{q_3^2+\tilde m_1^2}$. Similarly one finds that
$f_2(q)=0$ for $p_0-\sqrt{q_3^2+\tilde m_2^2}<q_0$ (where $\tilde
m_2$ has a meaning analogous to that of $\tilde m_1$, but for the
states with the momentum $-q+\frac{1}{2}(p+p')$).

As a result, due to the spectral condition (\ref{spectr}),
$f(q)=0$ in the region outside the hyperbola
\be\label{4.11'} p_0-\sqrt{q_3^2+\tilde
m_2^2}<q_0<-p_0+\sqrt{q_3^2+\tilde m_1^2}\ . \ee

To derive the \JLD representation, further we consider the
6-dimensional space-time with the Minkowskian metric
$(+,-,-,-,-,-)$. On this space, we define the vector
$z=(x_0,x_1,x_2,x_3,y_1,y_2)$. For practical purposes we introduce
also the notations for the 2-dimensional vector $\tilde x =
(x_0,x_3)$ and the 4-dimensional vector $\tilde z=(z_0, z_3,
z_4,z_5)\equiv(x_0, x_3, y_1,y_2)$. On the 6-dimensional space we
define the function
\be\label{F} F(z)=f(x)\delta(\tilde x^2-y^2)=f(x)\delta(\tilde
z^2), \ee
depending on all six coordinates.

When the causality condition (\ref{causal}) is fulfilled, i.e. for
the physical region, $f(x)$ and $F(z)$ determine each other, since
\begin{equation}\label{int}
\int dy_1dy_2 F(z)=f(x)\theta(\tilde x^2) =\left
\{\begin{tabular}{lll}
$f(x)$ & for & $\tilde x^2>0\ ,$ \\
0 &
for & $\tilde x^2<0\ .$\\
\end{tabular}
\right.
%&&f(x)\ \ \mbox{for}\ \ \tilde x^2>0\ ,\cr
%&&0\ \ \ \ \ \ \
%\mbox{for}\ \ \tilde x^2<0\ .
\end{equation}

The Fourier transform of $F(z)$,
\be F(r)=\int d^6ze^{izr}F(z)\ , \ee
can be expressed, using (\ref{F}) and (\ref{int}), as
\be\label{4.14} F(r)=\int d^4qD_1(r-\hat q)f(q)\ . \ee
Denoting the remaining 4-dimensional vector $\tilde r=(r_0, r_3,
r_4, r_5)$, we have
\be\label{D_1} D_1(r)=\int d^6ze^{izr}\delta(\tilde
z^2)=\frac{\delta(r_1)\delta(r_2)}{\tilde r^2}
=\delta(r_1)\delta(r_2)D_1(\tilde r)\ , \ee
with $D_1(\tilde r)=\frac{1}{\tilde r^2}$.

We define now the "subvector" of a 6-dimensional vector as $\hat
q=(q_0,q_1,q_2,q_3,0,0)$ and we find the relation between $F(\hat
q)$ and $f(q)$ in view of the causality condition (\ref{causal}):
\be\label{4.14'} F(\hat q)=\int d^4x f(x)\theta(\tilde
x^2)e^{iqx}=f(q)\ . \ee
$D_1(\tilde r)$ satisfies the 4-dimensional wave-equation:
\be \Box _4 D_1(\tilde r)=0\ , \ee
where the d'Alembertian is defined with respect to the coordinates
$r_0, r_3, r_4, r_5$. Then, due to (\ref{4.14}), it follows that
$F(r)$ satisfies the same equation,
\be\label{difeq} \Box _4F(r)=0\ . \ee
It is crucial to note that $F(r)$ depends on all six variables $r_0,
...r_5$:
$$
F(r)=\int d^4q f(q)D_1(\tilde r-\tilde
q)\delta(r_1-q_1)\delta(r_2-q_2)\ ,
%=\int d\tilde qf(\tilde q, r_1,r_2)D_1(\tilde r-\tilde q),
$$
where $\tilde q=(q_0, q_3, 0, 0)$.

The solution of (\ref{difeq}) can be written in the form
\cite{Vlad}:
$$
F(r')=\int d^3\Sigma_\alpha\int\int d r_1
dr_2\left[F(r)\frac{\partial D(\tilde r-\tilde r')}{\partial
\tilde r_\alpha} -D(\tilde r-\tilde r')\frac{\partial
F(r)}{\partial \tilde r_\alpha}\right]\delta(r_1)\delta(r_2)\ ,
$$
where $D(\tilde r)$ satisfies the homogeneous differential
equation $\Box _4D(\tilde r)=0$, with the initial conditions
$$
D(\tilde r)|_{r_0=0}=0 \ \ \ \mbox{and}\ \ \ \frac{\partial
D}{\partial r_0}(\tilde r)|_{r_0=0}=\prod_{i=1}^3\delta(r_i)\ .
$$
The first condition implies that $D(\tilde r)$ is an odd function,
with the result that:
\be\label{Dr} D(\tilde r)=\int d^4ze^{-i\tilde z\tilde
r}\epsilon(z_0)\delta(\tilde z^2)=\epsilon(r_0)\delta(\tilde r^2).
\ee
We note here that the surface $\Sigma$ is 3-dimensional and not
5-dimensional as it is in the commutative case with light-cone
causality condition. Now we can express $f(q)$ using (\ref{4.14'})
as:
\bn f(q)=F(\hat q)=\int d r_1 d r_2 \delta
(r_1-q_1)\delta(r_2-q_2)\cr \times\int d^3\Sigma_\alpha
[F(r)\frac{\partial D(\tilde r-\tilde q)}{\partial \tilde
r_\alpha} -D(\tilde r-\tilde q)\frac{\partial F(r)}{\partial
\tilde r_\alpha}]\ . \en
Due to the arbitrariness of the surface $\Sigma$, one can reduce
the integration over $r_4$ and $r_5$, using the cylindrical
symmetry, to the integral over $\kappa^2=r_4^2+r_5^2$.
Subsequently we change the notation of variables $r_i$ to $u_i$
and use the explicit form of $D(\tilde r)$ from (\ref{Dr}) to
obtain:
\begin{eqnarray}\label{interm}
&&f(q)=\int d u_1 d u_2\delta(u_1-q_1)\delta(u_2-q_2)\int
d^1\Sigma_j d\kappa^2\cr
&&\times\{F(u,\kappa^2)\frac{\partial}{\partial \tilde
u_j}\left[\epsilon(u_0-q_0)\delta((\tilde u-\tilde
q)^2-\kappa^2)\right]\cr &&-\epsilon(u_0-q_0)\delta((\tilde
u-\tilde q)^2-\kappa^2)\frac{\partial
F(u,\kappa^2)}{\partial\tilde u_j}\}\ .
\end{eqnarray}

Using the standard mathematical procedure \cite{Vlad} for
performing the integration in (\ref{interm}), we obtain the \JLD
representation in NC QFT, satisfying the light-wedge causality
condition (\ref{causal}):
\bn\label{4.18} f(q)&=&\int d^4u d\kappa^2\epsilon
(q_0-u_0)\delta[(q_0-u_0)^2-(q_3-u_3)^2-\kappa^2]\cr
&\times&\delta(q_1-u_1)\delta(q_2-u_2) \phi(u,\kappa^2)\ , \en
where $\phi(u,\kappa^2)=-\frac{\partial
F(u,\kappa^2)}{\partial\tilde u_0}$.

Equivalently, denoting $\tilde u=(u_0,u_3)$, (\ref{4.18}) can be
written as:

\be\label{4.18'} f(q)=\int d^2\tilde u d\kappa^2\epsilon
(q_0-u_0)\delta[(\tilde q-\tilde u)^2-\kappa^2] \phi(\tilde
u,q_1,q_2, \kappa^2)\ . \ee

The function $\phi(\tilde u, q_1,q_2, \kappa^2)$ is an arbitrary
function, except that the requirement of spectral condition
determines a domain in which $\phi(\tilde u,q_1,q_2, \kappa^2)=0$.
This domain is outside the region where the $\delta$ function in
(\ref{4.18'}) vanishes, i.e.
\be\label{4.19} (\tilde q-\tilde u)^2-\kappa^2=0\ , \ee
but with $\tilde q$ in the region given by (\ref{4.11'}), where
$f(q)=0$. Putting together (\ref{4.19}) and (\ref{4.11'}), we
obtain the domain out of which $\phi(\tilde u,q_1,q_2,
\kappa^2)=0$:
\begin{eqnarray}
&&a)\ \frac{1}{2}(\tilde p+\tilde p')\pm \tilde u\  \  \mbox{are in the forward light-wedge (cf. (\ref{spectr}))};\\
&&b)\ \kappa\geq max\left\{0, \tilde m_1-\sqrt{\left(\frac{\tilde
p+\tilde p'}{2}+\tilde u\right)^2}, \tilde
m_2-\sqrt{\left(\frac{\tilde p+\tilde p'}{2}-\tilde
u\right)^2}\right\}\ .\nonumber
\end{eqnarray}

For the purpose of expressing the scattering amplitude, we
actually need the Fourier transform $f_R(q)$ of the retarded
commutator,
\be\label{ret} f_R(x)=\theta(x_0)f(x)= \langle
p'|\theta(x_0)[j_1(\frac{x}{2}),j_2(-\frac{x}{2})]|p\rangle\ . \ee
Using  (\ref{ret}) and the Fourier transformation $f(x)=\int dq'
e^{-iq'x}f(q')$, we can express $f_R(q)$ as follows:
\bn\label{ret_q} f_R(q)&=&\int dx e^{iqx}f_R(x)=\int dx
e^{iqx}\theta(x_0)f(x)\cr &=& \int dq' f(q')\int dx
e^{i(q-q')x}\theta(x_0)\ . \en
Taking into account that
$$
\int dx_0 e^{i(q-q')x}\theta(x_0)=-i\frac{e^{i(\vec q-\vec q')\vec
x}}{q_0-q'_0}\ ,
$$
eq. (\ref{ret_q}) becomes:
$$ f_R(q)=i\int d q'_0 \frac{f(q'_0,\vec
q)}{q'_0-q_0}\ .$$

Now in the above formula we introduce the Jost-Lehmann-Dyson
representation (\ref{4.18'}), with the result:
\be\label{f_ret} f_R(q)=i\int \frac{d q'_0} {q'_0-q_0}\int
d^2\tilde u d\kappa^2\epsilon
(q'_0-u_0)\delta[(q'_0-u_0)^2-(q_3-u_3)-\kappa^2] \phi(\tilde
u,q_1,q_2, \kappa^2)\ .\ee
In (\ref{f_ret}) one can integrate over $q'_0$, using the known
formula of integration with a $\delta$-function, $\int
G(x)\delta(g(x))dx=\sum_i\frac{G(x_{0i})}{\frac{\partial
g}{\partial x}|_{x=x_{0i}}}$, where $x_{0i}$ are the simple roots
of the function $g(x)$. We identify in (\ref{f_ret})
$G(q'_0)=\frac{\epsilon(q'0-u_0)}{q'_0-q_0}$ and
$g(q'_0)=(q'_0-u_0)^2-(q_3-u_3)-\kappa^2$ (with the roots
$q'_0=u_0\pm\left[(q_3-u_3)^2+\kappa^2\right]^{1/2}$).

With these considerations, from (\ref{f_ret}) we obtain the NC
version of the \JLD representation for the retarded commutator:
\be\label{4.32} f_R(q)=\int d^2 \tilde u
d\kappa^2\frac{\phi(\tilde
u,q_1,q_2,\kappa^2)}{(q_0-u_0)^2-(q_3-u_3)^2-\kappa^2}\ . \ee
Compared to the usual \JLD representation,
\be\label{c_jld} f_R^{comm}(q)=\int d^4 u d\kappa^2\frac{\phi(
u,\kappa^2)}{(q_0-u_0)^2-(\vec q-\vec u)^2-\kappa^2}\ , \ee
the expression (\ref{4.32}) is essentially different in the sense
that the arbitrary function $\phi$ now depends on $q_1$ and $q_2$.
This feature will have further crucial implications in the
discussion of analyticity of the scattering amplitude in
$\cos\Theta$.

\subsection{(Non-)Analyticity of the scattering amplitude in $\cos\Theta$}

In the center-of-mass system (CMS) and in a set in which the
incoming particles are along the vector $\vec\beta=
(0,0,\theta)$\footnote{The 'magnetic' vector $\vec\beta$ is
defined as $\beta_i=\frac{1}{2}\epsilon_{ijk}\theta_{jk}$. The
terminology stems from the antisymmetric background field
$B_{\mu\nu}$ (analogous to $F_{\mu\nu}$ in QED), which gives rise
to noncommutativity in string theory, with $\theta_{\mu\nu}$
essentially proportional to $B_{\mu\nu}$ (see, e.g., \cite{SW}).},
the scattering amplitude in NC QFT depends still on only two
variables, the CM energy $E$ and the cosine of the scattering
angle, $\cos\Theta$ (for a discussion about the number of
variables in the scattering amplitude for a general type of
noncommutativity, see \cite{CMT}).

In terms of the \JLD representation, the scattering amplitude is
written as (cf. \cite{Schweber} for commutative case):
\be\label{scat} M(E,\cos\Theta)=i\int d^2\tilde u
d\kappa^2\frac{\phi(\tilde u, \kappa^2, k+p, (k'-p')_{1,2})}
{\left[\frac{1}{2}(\tilde k'-\tilde p')+\tilde
u\right]^2-\kappa^2}\ , \ee
where $\phi(\tilde u, \kappa^2,...)$ is a function of its $O(1,1)$-
and $SO(2)$-invariant variables: $u_0^2-u_3^2$,
$(k_0+p_0)^2-(k_3-p_3)^2$, $(k_1+p_1)^2+(k_2+p_2)^2$,
$(k'_1-p'_1)^2+(k'_2-p'_2)^2$,... The function $\phi$ is zero in a
certain domain, determined by the causal and spectral conditions,
but otherwise arbitrary.

For the discussion of analyticity of $M(E, \cos\Theta)$ in
$\cos\Theta$, it is of crucial importance that all dependence on
$\cos\Theta$ be contained in the denominator of (\ref{scat}). But,
since the {\it arbitrary} function $\phi$ depends now on
$(k'-p')_{1,2}$, it also depends on $\cos\Theta$. This makes
impossible the mere consideration of any analyticity property of
the scattering amplitude in $\cos\Theta$.

Since the \JLD representation reflects the effect of the causal and
spectral axioms, we notice that the hypotheses (\ref{causal}) and
(\ref{spectr}) used for the present derivation allow for a much
larger physical region, by not at all taking into account the effect
of the NC coordinates $x_1$ and $x_2$. One might wonder now whether
in the above derivation there is any condition which could be
subject to challenge. In that case there might also appear the
possibility that an analyticity domain can be obtained, leading to
some high-energy upper bound on the scattering amplitude.

\section{Causality in NC QFT}

\subsection{Causality and symmetry in NC QFT}

In the following, we shall challenge the causality condition
(\ref{causal})
\be\label{caus} f(x)=0\ , \,\ \ \mbox{for}\ \  \tilde x^2\equiv
x_0^2-x_3^2<0\ , \ee
which takes into account {\it only} the variables connected with the
$O(1,1)$ symmetry.

This causality condition is suitable in the case when the
nonlocality in the NC variables $x_1$ and $x_2$ is {\it infinite}.
The fact that in the causality condition (\ref{caus}) the
coordinates $x_1$ and $x_2$ do not enter means that the propagation
of a signal in this plane is instantaneous: {\it no matter how far
apart in the \nc coordinates two events are}, the allowed region for
correlation is given by only the condition $x_0^2-x_3^2>0$, which
involves the propagation of a signal only in the $x_3$-direction,
while the time for the propagation along $x_1$- and $x_2$-directions
is totally ignored.
%
%$$
%[x_1,x_2]=i\theta\ \ \Rightarrow\ \ \Delta x_1\Delta x_2\geq
%\frac{\theta}{2}\ \ \Rightarrow\ \ (\Delta x_1)^2+(\Delta
%x_2)^2\geq \theta\ .
%$$

Recall that we are using an axiomatic approach, in whose commutative
counterpart the assumption of locality was a postulate. In our
noncommutative case, the postulate of locality has to be replaced by
a postulate prescribing the scale of nonlocality. Postulating that
the scale of nonlocality in $x_1$ and $x_2$ is $l\sim\sqrt\theta$,
then the propagation of the interaction in the \nc coordinates is
instantaneous {\it only within this distance} $l$. It follows then
that two events are correlated, i.e. $f(x)\neq 0$, when
$x_1^2+x_2^2\leq l^2$ (where $x_1^2+x_2^2$ is the distance in the NC
plane with $SO(2)$ symmetry), provided also that $x^2_0-x^2_3\geq0$
(the events are time-like separated in the sense of $O(1,1)$).
Adding the two conditions, we obtain that
\be\label{nonloc} f(x)\neq 0\ , \ \mbox{for}\ \
x_0^2-x_3^2-(x_1^2+x_2^2-l^2)\geq 0\ .\ee
The negation of condition (\ref{nonloc}) leads to the conclusion
that the locality condition should indeed be given by:
$$
f(x)=0\ ,\ \ \ \mbox{for}\ \ \ \tilde x^2-(x_1^2+x_2^2-l^2)\equiv
x_0^2-x_3^2-(x_1^2+x_2^2-l^2)<0\ ,
$$
or, equivalently,
\be\label{newcaus} f(x)=0\ ,\ \mbox{for}\
x_0^2-x_3^2-(x_1^2+x_2^2)<-l^2\ , \ee
where $l^2$ is a constant proportional to the NC parameter $\theta$.
When $l^2\rightarrow 0$, (\ref{newcaus}) becomes the usual locality
condition.

When $x_1^2+x_2^2>l^2$, for the propagation of a signal only the
difference $x_1^2+x_2^2-l^2$ is time-consuming and thus in the
locality condition it is the quantity
$x_0^2-x_3^2-(x_1^2+x_2^2-l^2)$ which will occur. Therefore, we
shall have a again the locality condition of the form
(\ref{newcaus}).

Since there is no noncommutativity in the momentum space, the
spectral condition will read now as
\be\label{newspectr} p_0^2-p_1^2-p_2^2-p_3^2\geq 0,\ \ \ p_0>0\ .\ee

At this point we recall that the maximal symmetry of a NC QFT with
$\theta_{\mu\nu}$ a constant matrix is not the classical
$O(1,1)\times SO(2)$ symmetry, but a quantum symmetry, namely the
twisted \P\ symmetry \cite{CKT}, whose representation content is
identical to the usual \P\ symmetry. Moreover, the usual space-time
interval $x^2=x_0^2-x_1^2-x_2^2-x_3^2$ is invariant under the
twisted \P\ algebra, as well as the scale of nonlocality $l$, since
the latter is expressed in terms of the twisted \P-invariant
$\theta$. Consequently, (\ref{newcaus}) (\ref{newspectr}) are
compatible with the twisted \P\ algebra.

In fact, the consideration of nonlocal theories of the type
(\ref{newcaus}), (\ref{newspectr}) was initiated by Wightman
\cite{nonloc}. It was proven later \cite{Vlad, petrina, Wight} (see
also \cite{BLT}) that, indeed, in a quantum field theory which
satisfies the translational invariance and the spectral axiom
(\ref{newspectr}), the nonlocal commutativity
$$
[j_1(\frac{x}{2}),j_2(-\frac{x}{2})]=0\ ,\ \ \ \mbox{for}\ \ \
x_0^2-x_1^2-x_2^2-x_3^2<-l^2
$$
implies the local commutativity
\be\label{comm} [j_1(\frac{x}{2}),j_2(-\frac{x}{2})]=0\ ,\ \ \
\mbox{for}\ \ \ x_0^2-x_1^2-x_2^2-x_3^2<0\ . \ee

This powerful theorem (stating the "global nature of local
commutativity"), which does not require standard Lorentz invariance,
but only translational invariance, can be applied in the \nc case
with postulated finite nonlocality, with the conclusion that the
causality properties of a QFT with space-space \ncy are physically
identical to those of the corresponding commutative QFT.

It is then obvious that the Jost-Lehmann-Dyson representation
(\ref{c_jld}) obtained in the commutative case holds also on the NC
space for any orientation of the vector $\vec\beta$. Consequently,
the NC two-particle$\to$two-particle scattering amplitude will have
the same form as in the commutative case:
\be M(E,\cos\Theta)=i\int d^4 u d\kappa^2\frac{\phi( u, \kappa^2,
k+p)} {\left[\frac{1}{2}( k'- p')+ u\right]^2-\kappa^2}\ .\ee
This leads to the analyticity of the NC scattering amplitude in
$\cos\Theta$ in the analog of the Lehmann ellipse, which behaves
at high energies $E$ the same way as in the commutative case, i.e.
with the semi-major axis as
\be y_L=(\cos\Theta)_{max}=1 +\frac{const}{E^4}\ . \ee

\subsubsection {Enlargement of the domain of analyticity in
$\cos\Theta$ and use of unitarity. Martin's ellipse}

Two more ingredients are needed in order to enlarge the domain of
analyticity in $\cos\Theta$ to the Martin's ellipse and to obtain
the Froissart-Martin bound: the dispersion relations and the
unitarity constraint on the partial-wave amplitudes \cite{review}.

When using the causality condition (\ref{causal}), the forward
dispersion relation cannot be obtained in NC theory with general
direction of the $\vec\beta$-vector \cite{Liao}. However, the
conclusion to which we arrived by imposing the nonlocal
commutativity condition (\ref{newcaus}) and reducing it to the local
commutativity (\ref{comm}) leads straightforwardly to the usual
forward dispersion relation also in the NC case with a general
$\vec\beta$ direction.

As for the unitarity constraint on the partial wave amplitudes, the
problem has been investigated in \cite{CMT}, for a general case of
noncommutativity $\theta_{\mu\nu}$, $\theta_{0i}\neq 0$. For
space-space noncommutativity ($\theta_{0i}= 0$), the scattering
amplitude depends, besides the center-of-mass energy, $E$, on three
angular variables. In a system were we take the incoming momentum
$\vec p$ in the $z$-direction, these variables are the polar angles
of the outgoing particle momentum, $\Theta$ and $\phi$, and the
angle $\alpha$ between the vector $\vec\beta$ and the incoming
momentum. The partial-wave expansion in this case reads:
\be\label{pwe} A(E,
\Theta,\phi,\alpha)=\sum\limits_{l,l',m}(2l'+1)a_{ll'm}(E)Y_{lm}(\Theta,\phi)P_{l'}(\cos\alpha)\
, \ee
where $Y_{lm}$ are the spherical harmonics and $P_{l'}$ are the
Legendre polynomials.

Imposing the unitarity condition directly on (\ref{pwe}) or using
the general formulas given in \cite{CMT}, it can be shown that a
simple unitarity constraint which involves single partial-wave
amplitudes one at a time can be obtained only in a setting where the
incoming momentum is orthogonal to the NC plane (equivalently it is
parallel to the vector $\vec\beta$). In this case the amplitude
depends only on one angle, $\Theta$, and the unitarity constraint is
reduced to the well-known one of the commutative case, i.e.
\be Im\ a_l(E)\geq |a_l(E)|^2\ . \ee

For this particular setting, $\vec p \parallel\vec \beta$, it is
then straightforward, following the prescription developed for
commutative QFT, to enlarge the analyticity domain of scattering
amplitude to Martin's ellipse with the semi-major axis at high
energies as \be y_M=1 +\frac{const}{E^2}\ \ee and subsequently
obtain the NC analog of the Froissart-Martin bound on the total
cross-section, in the CMS and for $\vec p\parallel\vec \beta$:
\be\label{froissart} \sigma_{tot}(E)\leq c\ \ \ln^2 \frac{E}{E_0}\
. \ee

Thus, the unitarity constraint on the partial-wave amplitudes
distinguishes a particular setting ($\vec p\parallel\vec \beta$) in
which the Lehmann's ellipse can be enlarged to the Martin's ellipse
and the Froissart-Martin bound can be obtained, with the assumption
of finite nonlocality. Nevertheless, this does not necessarily
exclude the possibility of obtaining a rigorous high-energy bound on
the cross-section for $\vec p \nparallel\vec\beta$, and the issue
deserves further investigation.

\subsection{Causality and nonlocality in NC QFT}

It was shown in the previous subsection that the violation of
Lorentz invariance in itself does not forbid the existence of an
analyticity domain of the scattering amplitude in $\cos\Theta$ and
the derivation of a high-energy bound, compatible with the twisted
\P\ symmetry.

However, for the derivation of the analog of the Froissart-Martin
bound the key ingredient was the assumption of finite nonlocality.
This issue deserves a more thorough investigation, in the light of
the Lagrangean models studied so far. We have to point out from the
very beginning that the Lagrangean models have been studied up to at
most two-loops and that no definite statement about the
renormalizability of NC quantum field theories in general has been
made so far.

It is well known that in NC QFT treated with the Weyl-Moyal
correspondence (i.e. with the usual product of fields replaced by
Moyal $\star$-product in the Lagrangean) the short distance (UV)
effects are related to the long-distance (topological) features of
the space-time. This fact was first noticed in \cite{CDP}, where it
was shown that noncommutativity leads to UV-regular theories when at
most one dimension of the space-time is noncompact. For the NC flat
space-time UV-regularity is not achieved, but instead the exotic
phenomenon of UV/IR mixing appears \cite{UVIR}. The physical meaning
of this mixing is that at quantum level, even very low-energy
processes receive contributions from high energy virtual particles.
The nonlocality is energy-dependent, and for virtual particles of
arbitrarily high energy, the nonlocality is arbitrarily large.

Another investigation leading to the same conclusion was performed
in the first paper dealing with the causality in NC QFT in the
Lagrangean approach \cite{causal1}. There it was shown, through the
study of a scattering process, that space-space NC $\phi^4$ in 2+1
dimensions is causal at macroscopical level. However the incident
particles should be viewed as extended rigid rods, of the size
$\theta p$, perpendicular to their momentum. In other words, the
noncommutativity introduces an energy-dependent scale of spatial
nonlocality $\theta p$.

Judging by the above-mentioned results obtained in specific NC
models up to one-loop level, the previous analysis of analyticity
and high-energy bounds in axiomatic NC QFT becomes inconclusive. It
appears that the finite nonlocality condition (\ref{newcaus}) is
solely a conjecture, but only based on this conjecture one can
derive rigorously the analyticity properties and high-energy bounds
on the scattering amplitude (see Section 3). We should recall,
however, that the infinite nonlocality in NC QFT has been found up
to one-loop level and there is no indication that the infinite
nonlocality is not an artifact of perturbation theory.

Nevertheless, in the case of a {\it compact} \nc\ space-time, the NC
QFT is finite, i.e. there are no UV divergences \cite{CDP},
consequently no UV/IR mixing, and the range of nonlocality is
finite. For such NC QFT the finite nonlocality is no more a
conjecture and one may reconsider the rigorous axiomatic derivation
of the analyticity and high-energy bounds along the lines of Section
3.1.

\section{Conclusion and discussions}

In this paper we have tackled the problem of high energy bounds on
the two-particle$\to$two-particle scattering amplitude in NC QFT.
The key issue in the analysis proved to be the scale of nonlocality
of the quantum field theory on NC space-time.

We have found that, assuming infinite nonlocality and using the
causal and spectral conditions (\ref{causal}) and (\ref{spectr})
proposed in \cite{LAG} for NC theories with  $O(1,1)\times SO(2)$
symmetry, a new form of the Jost-Lehmann-Dyson representation
(\ref{4.32}) is obtained, which does not permit to draw any
conclusion about the analyticity of the scattering amplitude
(\ref{scat}) in $\cos\Theta$. Therefore the derivation of
high-energy bounds on the scattering amplitude is impossible.

However, by postulating that the nonlocality in the noncommuting
coordinates is finite, we were lead to imposing a new causality
condition (\ref{newcaus}), which accounts for the finitness of the
range of nonlocality and prevents the {\it instantaneous propagation
of signals} in the {\it entire} noncommutative plane $(x_1,x_2)$. We
proved that the new causality condition, compatible also with the
twisted \P\ symmetry, is formally identical to the one corresponding
to the commutative case (\ref{comm}), using the
Wightman-Vladimirov-Petrina theorem.

Thus, with the assumption of finite nonlocality, the scattering
amplitude in NC QFT is proved to be analytical in $\cos\Theta$ in
the Lehmann ellipse, just as in the commutative case; moreover,
dispersion relations can be written on the same basis as in
commutative QFT. Finally, based on the unitarity constraint on the
partial-wave amplitudes in NC QFT, we can conclude that, for
theories with space-space noncommutativity ($\theta_{0i}=0$), the
total cross-section is subject to an upper bound (\ref{froissart})
identical to the Froissart-Martin bound in its high-energy
behaviour, when the incoming particle momentum $\vec p$ is
orthogonal to the NC plane.

Though the perturbative studies performed so far (up to one loop)
indicate an infinite range of nonlocality as more plausible, it is
not yet clear whether this is a mere artifact of the perturbation
theory or not. Therefore a clear-cut conclusion about the existence
of high-energy bounds in NC QFT cannot be drawn, unless the question
of the scale of nonlocality is elucidated. In perturbative terms,
this is equivalent to the standing problem of UV/IR mixing. However,
for compact noncommutative spaces, where the range of nonlocality is
finite and the NC QFT models do not exhibit UV divergences, we trust
that an analog of the Froissart-Martin bound holds.

{\it Note added:} Recently, the validity of the Froissart-Martin
bound in NC QFT has been studied based on the AdS/CFT correspondence
\cite{Kumar}. The original idea appeared in \cite{Giddings}, where
the AdS/CFT correspondence was used to infer the Froissart-Martin
bound in high-energy QCD scattering. According to \cite{Kumar}, the
Froissart-Martin bound holds as well in NC QFT. This might look as
contradicting the results of the present paper. However, in
\cite{Kumar} the Froissart-Martin bound was derived in a specific
scalar field model, perturbatively and essentially by using an IR
cutoff brane. It turns out that in the considered toy model the
Froissart-Martin bound is saturated in both the commutative and
noncommutative directions, however the size of the cross-section is
smaller in the commutative directions than in the noncommutative
ones, with a ratio which depends only on the noncommutative
parameter and the IR cutoff. This strongly suggests that the IR
cutoff actually acts as a restriction on the range of nonlocality to
a finite region in the noncommutative plane.

\vskip 0.3cm {\Large\bf{Acknowledgements}}

We are grateful to C. Montonen, K. Nishijima, P. Pre\v{s}najder and
especially to M. Chaichian for useful discussions and remarks.

The financial support of the Academy of Finland under the Projects
No. 54023 and 104368 is acknowledged.

%\end{multicols}
\end{document}